\documentclass[preprint,amsmath,amssymb,superscriptaddress,aps,prl]{revtex4-1}
\usepackage{subfigure}
\usepackage{graphicx}
\usepackage{bm}
\usepackage{color}
\usepackage{times}
\usepackage{ulem}
\begin{document}
\title{Quasiparticle Interference of Surface States in Type-II Weyl Semimetal WTe$_2$}
\author{Wenhan Zhang}
\affiliation{Department of Physics and Astronomy, Rutgers University, Piscataway, New Jersey 08854, USA}
\author{Quansheng Wu}
\affiliation{Theoretical Physics and Station Q Zurich, ETH Zurich, 8093 Zurich, Switzerland}
\author{Lunyong Zhang}
\affiliation{Laboratory for Pohang Emergent Materials \& Max Plank POSTECH Center for Complex Phase Materials, Max Planck POSTECH/Korea Research Initiative, Pohang 790-784, Korea}
\author{Sang-Wook Cheong}
\affiliation{Department of Physics and Astronomy, Rutgers University, Piscataway, New Jersey 08854, USA}
\affiliation{Rutgers Center for Emergent Materials, Rutgers University, Piscataway, New Jersey 08854, USA}
\author{Alexey A. Soluyanov}
\affiliation{Theoretical Physics and Station Q Zurich, ETH Zurich, 8093 Zurich, Switzerland}
\affiliation{Department of Physics, St. Petersburg State University, St. Petersburg, 199034, Russia}
\author{Weida Wu}
\email{Corresponding Author.\newline wdwu@physics.rutgers.edu}
\affiliation{Department of Physics and Astronomy, Rutgers University, Piscataway, New Jersey 08854, USA}

\begin{abstract}
Topological Weyl semimetal (TWS) is a metal, where low energy excitations behave like Weyl fermions of high-energy physics. It was recently shown that due to the lower symmetry of condensed matter systems, they can realize two distinct types of Weyl fermions. The type-I Weyl fermion in a metal is formed by a linear crossing of two bands at a point in the crystalline momentum space - Brillouin zone (BZ). The second type TWSs host type-II Weyl points appearing at the touching points of electron and hole pockets, which is a result of tilted linear dispersion. The type-II TWS was predicted to exist in several compounds, including WTe$_2$.  Several ARPES studies of WTe$_2$ were reported so far, having contradictory conclusions on the topological nature of observed Fermi arcs. In this work, we report the results of spectroscopic imaging with a scanning tunneling microscope and first principle calculations, establishing clear quasiparticle interference features of the surface states of WTe$_2$. Our work provides a strong evidence for surface state scattering. Although the surface Fermi arcs are clearly observed, it is still difficult to prove the existence of predicted Type-II Weyl points in the bulk.
\end{abstract}

\maketitle
Weyl fermion was first predicted in particle physics in the beginning of the quantum era~\cite{weyl1929elektron}. Although an example of a Weyl fermion is still unknown in particle physics, in condensed matter, it was theoretically proposed to emerge in topologically non-trivial crystals~\cite{Wan2011,Burkov2011,Bulmash2014,Liu2014d}. The material hosting Weyl fermions is called topological Weyl semimetal (TWS). Weyl fermions in these materials always emerge in pairs of opposite chirality, being either a source or a sink of Berry curvature. Consequently, they can only be annihilated in pairs, being otherwise stable to weak translation-preserving perturbations~\cite{Nielsen1981,Nielsen1981b}. The first type of Weyl fermion was predicted and observed in TaAs family of compounds~\cite{weng2015a,Xu2015,Lv2015c,Lv2015b}. It is formed by a linear crossing of the valence band and conduction band in the Brilloum zone (BZ). Its low-energy excited states behave like Weyl fermions of standard quantum field theory. The Fermi surface in type-I TWSs, formed by Weyl points, is always closed. Interestingly, in the surface spectrum, this results in the appearance of open Fermi arcs, connecting the projections of opposite chirality Weyl points to the surface. Besides, TWSs supposedly provide a condensed matter realization of the chiral anomaly~\cite{Nielsen1983}. Shortly after the type-I Weyl fermions were realized, Ref.~\onlinecite{Soluyanov2015} proposed type-II TWSs, in which the linear dispersion is tilted so that Weyl points appear at the touching points of electron and hole pockets. Unlike type-I Weyl points, these Weyl points always have an open Fermi surface (when the Hamiltonian is linear in momentum), resulting in the unusual chiral anomaly~\cite{Xu2015c}. Type-II TWS was predicted in several transition metal dichalcogenides (TMD): WTe$_2$~\cite{Soluyanov2015}, MoTe$_2$~\cite{Sun2015,Qi2016} and Mo$_x$W$_{1-x}$Te$_2$~\cite{Chang2016a}.

TWSs present many interesting exotic properties, such as surface Fermi arcs~\cite{Wan2011,Xu2011a} and unconventional magneto-transport phenomena due to the chiral anomaly~\cite{Nielsen1983,Zyuzin2012,Son2013,Spivak2016,Hosur2013,Parameswaran2014,Burkov2015}. Among the interesting properties of TWSs, surface Fermi arc is a crucial property associated with the non-trivial topological nature of the bulk states. Therefore, visualizing such surface states, and proving their topological origin, is one of the major efforts in studies of TWS. Angle-resolved photoemission spectroscopy (ARPES) studies on both type-I (TaAs family)~\cite{Xu2015,Lv2015c,Lv2015b,Yang2015a} and type-II TWS (transition metal dichalcogenides)~\cite{Huang2016,Deng2016,Wu2016,Bruno2016a,Wang2016,Jiang2017} have provided evidence of surface states, but whether the appearance of these states is a result of type-II Weyl points in the bulk remained unanswered. STM is a powerful technique to characterize the surface states of TWS via quasiparticle interference (QPI) from spectroscopy measurments~\cite{Kourtis2015,Chang2016}. It can measure both the occupied and unoccupied states with excellent energy resolution, while APRES accesses only occupied states normally. This motivated scanning tunneling spectroscopy (STS) on WTe$_2$ to visualize the QPI due to surface states.

\begin{figure*}[ht]
\centering
\includegraphics[width=0.95\textwidth]{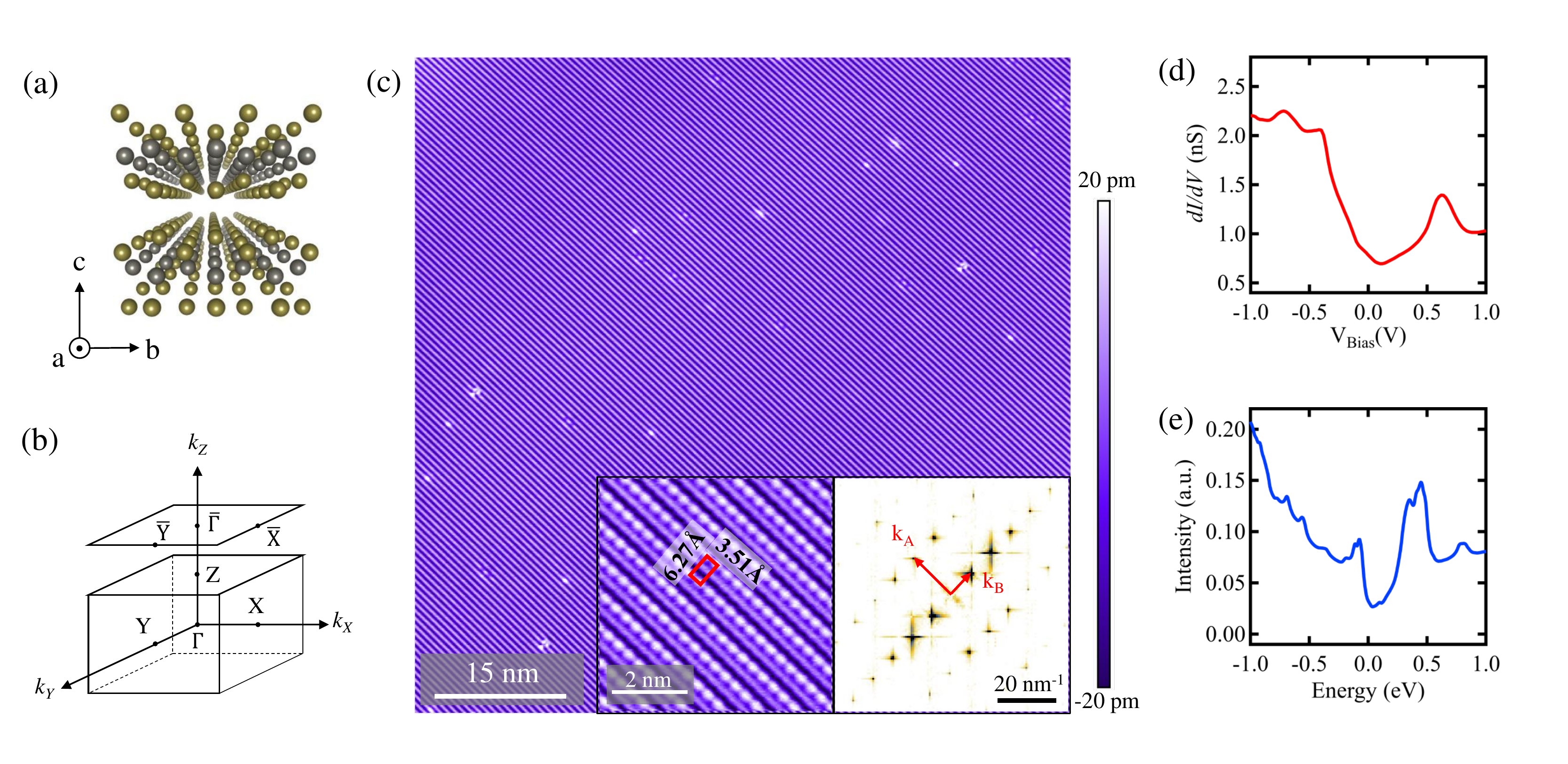}
\caption{(color online). (a) Crystal structure of WTe$_2$. (b) Schematics of bulk and surface Brillouin zones. (c) Large-scale topographic image of WTe$_2$. ($V_B= 1$~V, $I_T= 10$~pA) Left inset: Zoom-in topographic image ($V_B= 0.01$~V, $I_T= 2.4$~nA). Right inset: Fourier transform of the left inset image showing Bragg peaks corresponding to the atomic corrugation. (d) Average $dI/dV$ spetrum taken on WTe$_2$ surface. ($V_B= -1$~V, $I_T= 1$~nA) (e) Calculated total density of state.}
\label{fig1}
\end{figure*}

STS measurments have been performed on several TWS materials, such as TaAs~\cite{Inoue2016,Avraham2016,Sessi2017a}, NbP~\cite{Zheng2016} and MoTe$_2$~\cite{Deng2016}. However, despite several pior STM stuides on WTe$_2$~\cite{Li2016,Das2016}, the clear evidence of surface states of WTe$_2$ above $E_F$ is still absent. Due to coexistence of the electron and hole pockets with Weyl nodes near the Fermi energy, the bulk states also contribute significantly to the surface scattering, which complicates the identification of the surface states. Therefore, it is necessary to compare experimental QPI results with first principle calculations to differentiate QPI feature of surface states from that of bulk states.

In this work, we used STM/STS to directly visualize QPI patterns of surface states from two distinct surfaces of WTe$_2$ single crystals. For comparison, we also carried out density functional theory (DFT) calculations and obtained surface spectral weight maps as well as the corresponding spin-preserved joint density of state (JDOS) maps at various energies. The good agreement between DFT calculations and experiments confirms that the main QPI is from scattering between two surface Fermi arcs in the surface BZ. The solid evidence of surface states on WTe$_2$ will stimulate further investigation of topological nature of surface states in TWS materials.

High quality single crystals of WTe$_2$ were grown via iodine vapor transport method. Tungsten powder ($99.9\%$) and tellurium powders ($99.99\%$) were well-mixed and heated in an evacuated silica tube at $700~^{\circ}$C for 2 days; the synthesized product was then ground and heated at $750~^{\circ}$C for 2 days. The final pellet was ground into fine powders. Appropriate amount of powders and iodine were sealed in an evacuated silica tube and put in a two-zone furnace with a temperature gradient of $50~^{\circ}$C between $850~^{\circ}$C and $800~^{\circ}$C for 1 week.

STM/STS measurements were carried out at $4.5$~K in an Omicron LT-STM with base pressure of $1\times10^{-11}$~mbar. Electrochemically etched tungsten tip was characterized on single crystal Au(111) surface. To differentiate two polar surfaces, one piece of WTe$_2$ single-crystal was cut into two halves and one of them was flipped upside down. Then both samples were mounted on STM sample plates without changing their orientations. They were cleaved \textit{in-situ} in ultra-high vacuum at room temperature, then were immediately transferred into cold STM head for measurements. Since WTe$_2$ single crystals are always in single domain state, the cleaved surfaces of the flipped crystal is presumably opposite to the other one. The $dI/dV$ grid mapping measurements were performed to probe the QPI of the surface states. The setpoint is $V_B$= $-0.1$~V, $I$= 0.5~nA. At each point, a full $dI/dV$ spectrum was recorded by ramping $V_B$ from -0.1 to 0.1~V with feedback off. The standard lock-in technique was utilized with modulation frequency $f= 455$~Hz and modulation amplitude $V_{mod}= 20$~mV. The Fourier transform of $dI/dV$ maps are mirror symmetrized about $q_y$ axis and then smoothed with Gaussian function.

On the theoretical side, we performed electronic structure calculations using DFT as implemented in the Vienna Ab Initio simulation package (VASP)~\cite{Kresse1996} with projector augmented wave basis sets~\cite{paw} that included spin-orbit coupling. The PBE functional~\cite{pbe} was used in the exchange-correlation potential. A 16$\times$10$\times$4 $\Gamma$-centered k-point mesh was used for Brillouin zone sampling, and the energy cutoff was set to 450~eV. Then Wannier-based projected tight-binding models\,\cite{wannier90, Marzari97, Souza-Marzari02} capturing all the $s$ and $d$ states of W and $p$ states of Te were used to analyze the surface density of states. Surface spectra were calculated by the software package WannierTools~\cite{Wu2017}, which is based on the iterative Green's function mechanism~\cite{Sancho1985}. The spin-dependent joint density of states (JDOS) were calculated as
\begin{eqnarray}
J_s({\bf q})= \frac{1}{2}\sum_{{\bf k}}\sum_{i=0, 1,2,3}\rho_i({\bf k})\rho_i({\bf k+q})
\end{eqnarray}
where total spectral density $\rho_0({\bf k})= -\frac{1}{\pi}\text{Im}  [\text{Tr} G_s({\bf k},\epsilon_0)]$, $G_s({\bf k},\epsilon_0)$ is the surface Green's function at momentum k and energy $\epsilon_0$ relatively to Fermi energy, and the spin density $\rho_i({\bf k})= -\frac{1}{\pi}\text{Im}  [\text{Tr} [\sigma_i G_s({\bf k},\epsilon_0)]], i=1, 2, 3$ with $\sigma_{1,2,3}$ being the Pauli matrices of electron spin.

\begin{figure*}[ht]
\centering
\includegraphics[width=0.9\textwidth]{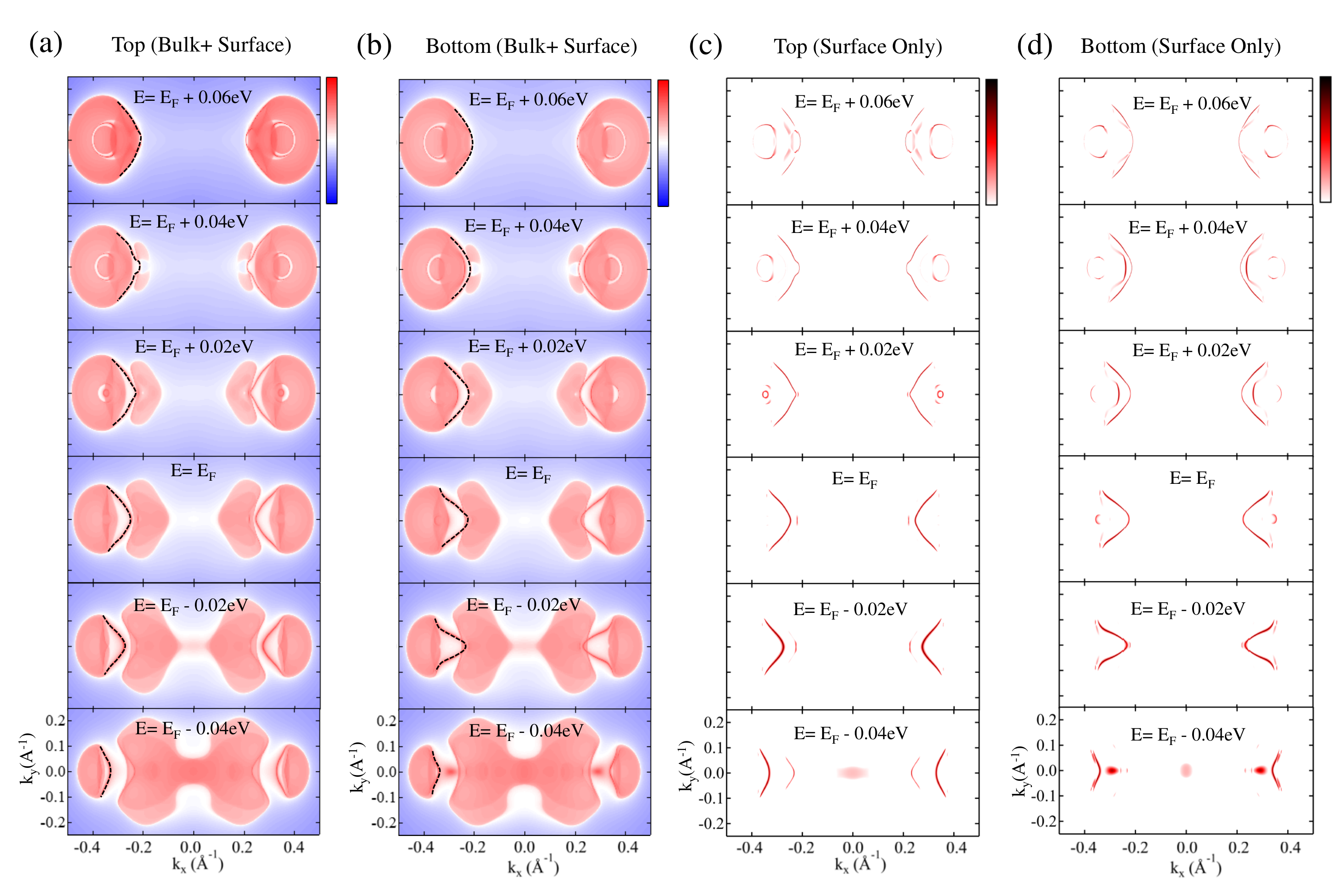}
\caption{(color online). Calculated spectral weight maps: the bulk states and surface states on topmost (a) and bottommost (b) Te layers of WTe$_2$; the surface states only on topmost (c) and bottommost (d) Te layers of WTe$_2$. The energy is between $E_{F}- 40$~meV and $E_{F}+ 60$~meV. The dashed lines indicate the surface Fermi arcs.}
\label{fig2}
\end{figure*}

Fig.~\ref{fig1}(a) and (b) display the schematic bulk atomic structure and its Brillouin zone of WTe$_2$. Due to the lattic distortion, WTe$_2$ has orthorhombic unit cell with the space group $Pmn2_1$~\cite{Ali2014,Soluyanov2015}. Correspondingly, the surface Te atoms distort and form chains along $a$ axis, as shown by large scale topography in Fig~\ref{fig1}(c). In the zoom-in topographic image [the left inset of Fig.~\ref{fig1}(c)], two inequivalent Te atomic chains are visible. The one with higher apparent height has better atomic resolution. The lattice constants estimated by the Bragg peaks in the right inset of Fig.~\ref{fig1}(c) are $a=3.51$~\AA\,and $b= 6.27$~\AA , which are consistent with the previous reports~\cite{Ali2014,Li2016}. The orthorhombic lattice structure was repeatedly observed on multiple pieces of WTe$_2$ samples within our STM orthogonal uncertainty ($< 2~^{\circ}$). The local density of states (LDOS) were measured by the $dI/dV$ spectrum [Fig.~\ref{fig1}(d)] showing a semi-metallic behavior. The calculated total DOS of bulk WTe$_2$ [Fig.~\ref{fig1}(e)] qualitatively agrees with the measurements.

Prior band structure calculations of WTe$_2$~\cite{Ali2014} reveal that both the valence band and conduction band cross the Fermi level, forming electron and hole pockets. This has been confirmed by ARPES studies~\cite{Pletikosic2014}. The calculations also predict that the Weyl points locate at around $50$~meV above $E_F$~\cite{Soluyanov2015}. Fig.~\ref{fig2} (a) and (b) plot the calculated spectral weight maps of bulk and surface states at various energies by projecting electronic states to the surface Te layers. The surface Fermi arcs are marked by the black dashed lines. The hole pockets locate closer to the $\Gamma$ point than the Fermi arcs. As the energy increase, they shrink and disappear. The electron pockets, on the other hand, locate farther from the $\Gamma$ point than the Fermi arcs. They grow larger at higher energy. In order to emphasize the surface states, projections from the bulk states have been removed. The spectral weigh maps of the surface states are display in Fig.~\ref{fig2} (c) and (d). Due to the broken inversion symmetry, the top and bottom surfaces of WTe$_2$ have inequivalent band structures. And this is confirmed by our calculated surface Fermi arcs on different surfaces. However, the dispersions of the surface Fermi arcs on these two surfaces are qualitatively the same. The surface Fermi arcs have non-trivial spin texture, which has been reported in Ref.~\onlinecite{Feng}. They are most visible from $E_{F}-40$~meV up to $E_{F}+60$~meV. As the energy increases, the arc-like surface states gradually moves toward $\Gamma$ point (the BZ center). They completely disappear above $E_F+60$~meV. It is in good agreement with constant energy contour results in APRES experiments~\cite{Bruno2016a,Wu2016,Wang2016}.

\begin{figure*}[ht]
\centering
\includegraphics[width=1\textwidth]{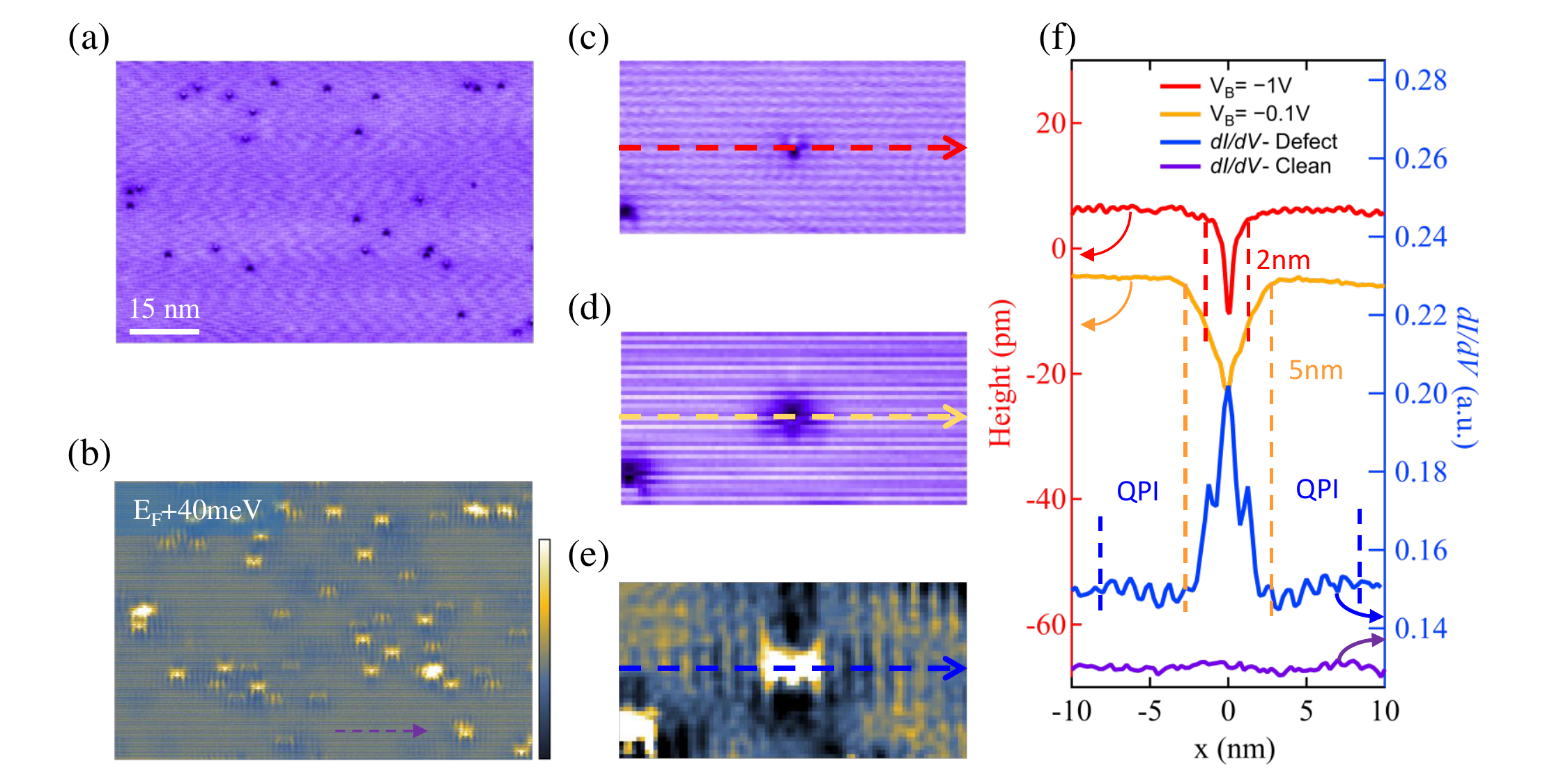}
\caption{(color online) (a) Topographic image where the $dI/dV$ grid mapping was performed. ($V_B= -1$~V, $I_T= 100$~pA) (b) The $dI/dV$ map at $E$= $E_F + 40$~meV in the same field of view as (a), showing quasiparticle scattering patterns around defects. (c) A zoom-in topographic image of an individual defect at $V_B= -1$~V. (d) A topographic image at $V_B= -0.1$~V in the same field of view as (c). (e) A $dI/dV$ map at $E$= $E_F + 40$~meV in the same field of view as (c). (f) Line profiles of the topographic heights(red and orange) and $dI/dV$ signals(blue). The positions where they were taken are marked by arrows in (c)-(e). The purple curve is the dI/dV line profile in defect-free area, as marked in (b).} 
\label{fig3}
\end{figure*}

With the guidance of calculated spectral weight maps of surface states, we performed spectroscopic grid mapping on two distinct surfaces of WTe$_2$ to investigate energy-dependent QPI patterns, which are the result of electrons being elastically scattered by defects. We locate a region on surface-1 with sufficient defect density, as shown in Fig.~\ref{fig3}(a). The fast scan axis is chosen to be parallel with the atomic chain ($a$ axis). The major defects appear as surface suppression, indicating they may be subsurface vacancies or anti-sites. Clear spatial scattering around defects was observed in the $dI/dV$ maps between $-100$~meV and $+100$~meV. Fig.~\ref{fig3}(b) displays a representative $dI/dV$ map at $+40$~meV. Note that there are more scattering centers in the $dI/dV$ map [Fig.~\ref{fig3}(b)] than the surface defects observed in the topographic image of Fig.~\ref{fig3}(a). They are probably defects underneath the surface Te layer. Two different patterns are commonly observed in the $dI/dV$ maps at various energies: one is the $dI/dV$ modulations localized on top of the defect cites (white spots at the center of Fig.~\ref{fig3}(e)); the other is the much weaker but extended standing waves around defects. Fig.~\ref{fig3}(f) displays the line profiles taken on the topographic images [Fig.~\ref{fig3}(c)(d)] and the $dI/dV$ map [Fig.~\ref{fig3}(e)] across an isolated defect. The length scale of that localized $dI/dV$ modulation is $l\sim$5~nm, the same as the defect size, indicating that it is a result of impurity potential or defect states. In contrast, the oscillating pattern spreads about 17~nm away from the defects with well defined spatial periodicity $\lambda\sim$1.1~nm along $a$ direction, suggesting that it is the QPI pattern. Thus, based on the different length scales, the QPI pattern is unambiguously separated with the $dI/dV$ modulations induced by the defect potential. The spectroscopic data of the other surface (not shown) present the same features.

\begin{figure*}[ht]
\centering
\includegraphics[width=0.8\textwidth]{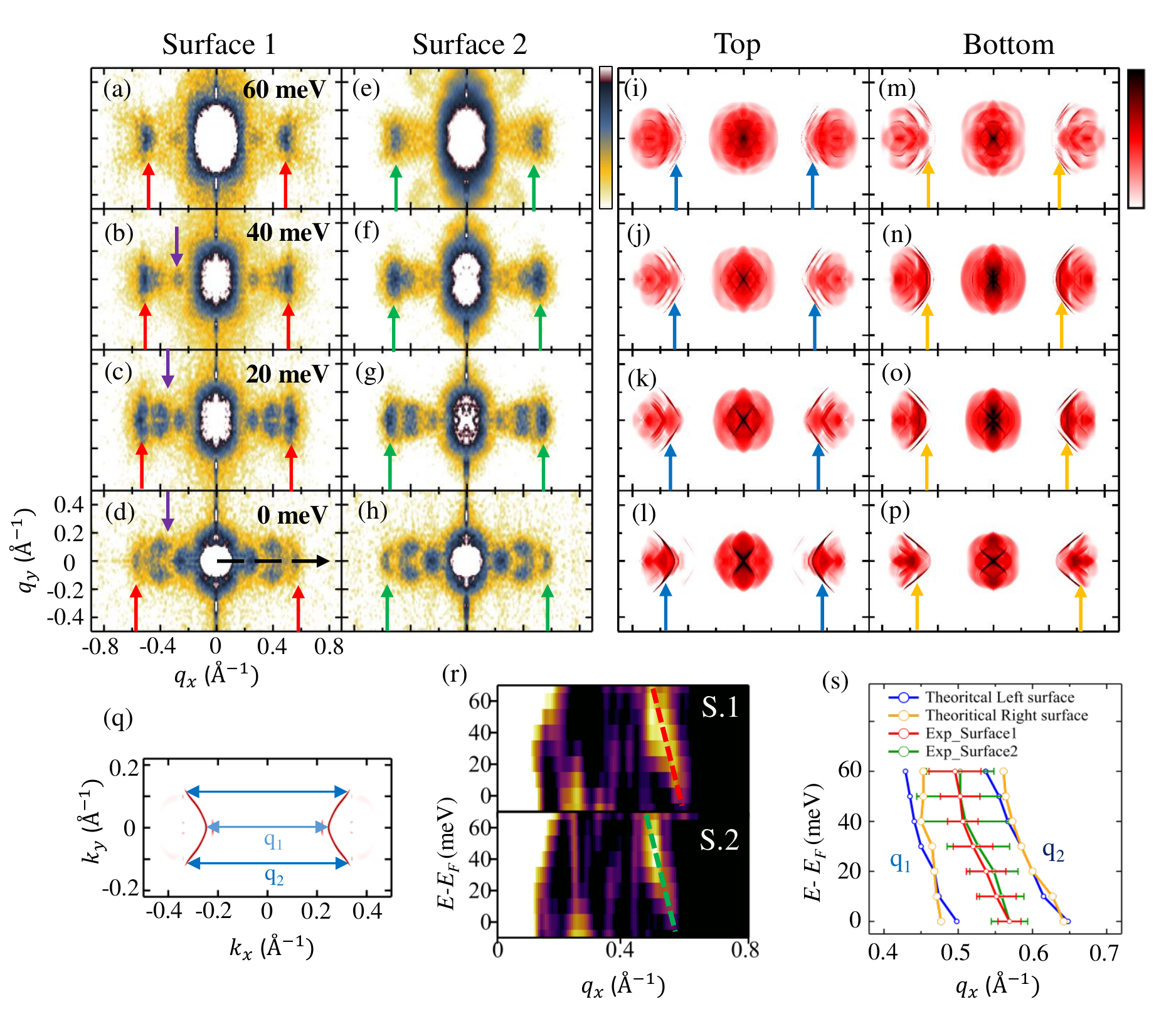}
\caption{(color online) (a)-(d) Fourier transforms of $dI/dV$ maps taken on surface 1 from $E_F$ to $E_F+ 60$~meV. (e)-(h) Fourier transforms of $dI/dV$ maps taken on surface 2 in the same energy range. The red arrows points to inter-arc scattering features. Blue arrows point to the features generated by scattering between bulk states. (i)-(p) Calculated spin-preserving JDOS maps of surface states of both the topmost and the bottommost surfaces in the same energy range. (q) Schematic of inter scattering between two Fermi arcs in the BZ. $q_1$ and $q_2$ represents the head-to-head and tail-to-tail scattering vectors. (r) The experimental E-Q dispersion along $q_x$ axis on two surfaces. The green dashed line marks the dispersion. (s) The dispersion of scattering vectors extracted from QPI patterns and comparison with calculated $q_1$ and $q_2$.}
\label{fig4}
\end{figure*}

Fourier transforms (FTs) of the $dI/dV$ maps in Fig.~\ref{fig4}(a)-(h) display the $q$ (scattering wave vector) maps of QPI in the surface BZ in the energy range from 0 up to $+60$~meV. Several non-trivial features were observed: red and green arrows point to a sharp pattern evolving towards $\Gamma$ point as the energies increases, and the purple arrows mark a non-dispersive pattern that fades out gradually and disappears when $E> E_F+ 40$~meV. The FT intensity around $\Gamma$ point is very high, but no sharp features are observed there. The sharp pattern locates at $q_x\sim$0.55~\AA$^{-1}$, equivalent to $\frac{2\pi}{\lambda}$, indicating its correspondence to the QPI shown in Fig.~\ref{fig3}(f). Its dispersive character is manifestly illustrated in Fig.~\ref{fig4}(r). As for the $dI/dV$ modulation localized at defect centers, the corresponding $q$ value is about 0.13~\AA$^{-1}$, which is much closer to the $\Gamma$ point than the observed QPI. To understand the origins of these features in the $q$ maps, the surface JDOS maps were calculated.

WTe$_2$ is a non-magnetic type-II TWS. The time reversal symmetry prevents scatterings between the states with opposite spins. This has been confirmed by STM studies of other non-magnetic TWSs~\cite{Inoue2016,Zheng2016,Deng2016,Zheng2016a}. Therefore, the spin-preserving JDOS calculations are necessary to compare to the FTs of the $dI/dV$ maps. Fig.~\ref{fig4}(i)-(p) show the images of spin-preserved JDOS of surface states at the same energies with Fig.~\ref{fig4}(a)-(h). The X-shaped feature at $\Gamma$ point mainly originates from the intra-arcs scattering of the Fermi arcs and the arc-like features on the left and right sides correspond to the inter-arc scattering. As the energy increases, the inter-arc scattering features marked by the blue and orange arrows gradually move towards $\Gamma$ point. This trend is consistent with the evolving feature marked by red and green arrows in the FTs of the $dI/dV$ maps in Fig.~\ref{fig4}(a)-(h), suggesting that the red- and green-arrow pattern originates from the surface state scattering.

To quantitatively compare the experimental QPI maps and calculated JDOS maps, the dispersions of the sharp pattern in Fig.~\ref{fig4}(a)-(h) are extracted and plotted together with the scattering vector obtained from DFT calculations, as shown in Fig.~\ref{fig4}(s). Here $q_1$ and $q_2$ represent the scattering vectors connecting the heads and tails of the two Fermi arcs respectively, as illustrated by Fig.~\ref{fig4}(q). With the presence of time reversal symmetry, the states at the apex of the two Fermi arcs ($k_y=0$) must have opposite spins, meaning the surface state scattering $q_1$ is suppressed. Yet scattering vectors between $q_1$ and $q_2$ are allowed. The experimentally observed scattering vectors indeed fall in the range between $q_1$ and $q_2$ at various energies. The good agreement shows that the sharp feature captured in QPI patterns originates from scattering between surface states of Fermi arcs. In addition, the pattern marked by purple arrows in Fig.~\ref{fig4}(a)-(h) does not associate with the surface state scattering. It may result from the scattering between bulk states. Within the experimental uncertainty, the two surfaces present essentially identical QPI features. It is also consistent with our DFT calculation results, that the two surfaces have qualitatively the same electronic structure.

It is still under debate whether the observed surface states are topological or trivial~\cite{Bruno2016a,Wu2016,Wang2016}. On one hand, small changes of lattice structure may change the topological nature of the material~\cite{Bruno2016a}. On the other hand, even in the topological phase, the trivial and topological surface states coexist and are in close proximity in $k$-space, so it is very difficult to distinguish them experimentally. Furthermore, a recent work reported the Rashba spin splitting effect on WTe$_2$ surface~\cite{Li2016}, suggesting that the trivial surface states may also be spin-polarized. This makes it more difficult to identify the topological surface states unambiguously.

In conclusion, we combined STM/STS measurements and first principle calculations to resolve the surface states on the (001) surface of WTe$_2$. The QPI patterns indicate the scattering of dispersive surface states. The calculated spin-dependent JDOS maps further confirm the existence of surface states on WTe$_2$. Our work provides evidence of the surface states on Type-II TWS WTe$_2$ and may inspire the subsequent research to figure out the topology of such surface states in semi-metallic TWSs.

We are grateful to J. Li, Z. Wang, and A. Bernevig for helpful discussions. The STM work at Rutgers was supported by NSF Grants No. DMR-1506618. The crystal growth work was partially supported by the NSF under Grant No. DMR-1629059. It was also supported by the Max Planck POSTECH/KOREA Research Initiative Program through National Foundation of Korea (NRF) funded by the Ministry of Science, ICT and Future Planning(No. 2016K1A4A4A01922028). The calculation work was supported by the European Research Council through the ERC Advanced Grant SIMCOFE, the Swiss National Science Foundation and through the National Competence Centers in Research MARVEL and QSIT. Q.W. and A.S. also acknowledge the support of Microsoft Research.

W.Z. and Q.W. contributed equally to this work.

%
\end{document}